\documentclass{cta-author}

\usepackage{amsmath,amssymb,amsfonts}
\usepackage{algorithmic}
\usepackage{graphicx,subfigure}
\usepackage{textcomp,mathrsfs}
\usepackage{xcolor,booktabs}
\usepackage{steinmetz,relsize}
\usepackage{enumitem,balance,lastpage}
\usepackage[colorinlistoftodos]{todonotes}
\usepackage{color,caption,soul}
\setlist{  
  listparindent=\parindent,
  parsep=0pt,itemsep=0pt,
}

{}
{}
{}

\begin{document}

\supertitle{Submission Template for IET Research Journal Papers}

\title{An Investigation into the Impacts of Deep Learning-based Re-sampling on Specific Emitter Identification Performance}

\author{\au{Mohamed K. M. Fadul$^{1}$}, \au{Donald R. Reising$^{1\corr}$}, \au{Lakmali P. Weerasena$^{2}$}}

\address{\add{1}{Electrical Engineering Department, University of Tennessee at Chattanooga, 735 Vine Street, Chattanooga, TN, U.S.A.}
\add{2}{Department of Mathematics, University of Tennessee at Chattanooga, 735 Vine Street, Chattanooga, TN, U.S.A.}
\email{donald-reising@utc.edu}}

\begin{abstract}
Increasing Internet of Things (IoT) deployments present a growing surface over which villainous actors can carry out attacks. This disturbing revelation is amplified by the fact that a majority of IoT devices use weak or no encryption at all. Specific Emitter Identification (SEI) is an approach intended to address this IoT security weakness. This work provides the first Deep Learning (DL) driven SEI approach that upsamples the signals after collection to improve performance while simultaneously reducing the hardware requirements of the IoT devices that collect them. DL-driven upsampling results in superior SEI performance versus two traditional upsampling approaches and a convolutional neural network only approach.
\end{abstract}

\maketitle

\section{Introduction%
\label{sec:introduction}}
It is anticipated that the number of deployed Internet of Things (IoT) devices will reach 75 billion by 2025~\cite{Statista_IoT_2019}. However, limited computational resources, high manufacturing costs, and scalability challenges associated with implementation and management of encryption has led most IoT devices ($\sim$70\%) to weak employments or to forego it altogether~\cite{Ray_CIC_2019}. This leaves the IoT devices and corresponding infrastructure exposed to abuse by nefarious actors~\cite{Wright_Killer_2019}. Specific Emitter Identification (SEI) is a passive, physical layer means to secure IoT devices and the associated infrastructure~\cite{KandahiThings2019,Reising_IoT_2020}. SEI exploits inherent, unique, and distinct features that are unintentionally imparted upon a radio's signals during their formation and transmission. These features are attributed to acceptable manufacturing discrepancies and differences present within the components, sub-systems, and systems that comprise the radio's Radio Frequency (RF) front-end. Although sufficient to discriminate radios of the same manufacture and model, SEI features do not interfere with a radio's normal operations. Recently, Deep Learning (DL) based SEI has been put forward, because it has been shown capable of learning discriminator features directly from the In-phase (I) and Quadrature (Q) values of collected, sampled signals~\cite{Restuccia_DeepRadioID_2019}. This eliminates the need for handcrafted SEI that requires expert knowledge to carefully select signal transformations, feature generation and selection techniques, and other practices that maximize performance. Despite its successes, SEI performance is greatly influenced by the receiver's capabilities such as sampling rate. Typically, SEI makes use of high sampling rates to ensure the nuances of a given radio's signal features are sufficiently captured. However, this is problematic when considering IoT devices, because they may be incapable of generating sampling rates high enough to capture these nuances. If this occurs, then SEI performance will be degraded and lead to radios being misidentified, which can result in network access being granted to 
unauthorized IoT devices and users.

This work proposes the novel use of a conditional Generative Adversarial Network (cGAN) to increase the sampling rate of signals collected at lower sampling rates prior to DL-based SEI~\cite{CGAN_Mizra}. The goal is to improve SEI performance without increasing the hardware requirements (e.g., memory, clock rate) of the IoT devices that collect the signals. The use of a cGAN permits generation of higher sampled signals from the learned SEI feature distributions associated with each of the radios being identified while eliminating the need to learn Markov Chain parameters~\cite{Goodfellow_NIPS2014}. The proposed cGAN-based upsampling approach is compared with two traditional interpolation techniques--described in~\cite{deboor}--that are each applied to the same set of IoT device signals. 

The remainder of this article is organized as follows: Sect.~\ref{sec:related_works} provides a brief summary of related works and describes how they differ from the work presented in this article; Sect.~\ref{sec:background} describes the signal of interest, the employed deep learning architectures, as well as descriptions of two, traditional interpolation techniques used to facilitate comparative assessment; Sect.~\ref{sec:methodology} outlines the process developed to generate the results in Sect.~\ref{sec:results}; and the conclusion is presented in Sect.~\ref{sec:conclusio}.
\section{Related Works
\label{sec:related_works}}
A key motivating factor for DL-based SEI is the elimination of feature-engineering or domain-specific, expert knowledge. DL-based SEI 
has been well investigated by multiple efforts~\cite{Baldini_interference_2019, Baldini_comparison_2019, Shea_Air_2019, wong_clustering_2018, Gihan_hardware_2019, Riyaz_CNN_2018, Merchant_DL_fingerprinting_2018, Restuccia_DeepRadioID_2019, Jafari,Guyue_Wifi_2019, Youssef_ML_2018,Pan_2019,Yu_WiMob_2019,Jian_Massive_study_2020,Robinson_Dilated_2020, Ding_IET_2018, Chen_IET_2020, Peng_TVT_2020, Yang_ICT_2021, Morin_SIP_2019, Behura_TCCN_2020, McGinthy_IoT_2019, Tang_2021, Liu_IoT_2021, Zha_ICT_2021, Gong_WCSP_2019, Ji_AUTEEE_2020, Gong_infosec_2020, Li_AIID_2021, Wang_Comm_2021, Qu_Sym_2021, Bassey_FMEC_2019, Wang_ICICSP_2021, Cun_ICCT_2021, Peng_CCC_2021, Shen_CIS_2018,Tyler_ICC_2022}. The work in \cite{Gong_WCSP_2019, Gong_infosec_2020} proposes semi-supervised SEI approaches based upon GANs. The authors of \cite{Gong_WCSP_2019} extract the the received RF signals' features using a representation network and then train a Triple-GAN network using the extracted features. Feedback learning assists the representation network to learn more discriminative SEI features while the representation network finds a better raw data representation to improve the performance of the Triple-GAN and SEI classification system.
The work in~\cite{Gong_infosec_2020} performs SEI using a Information maximized GAN (InfoGAN) and applies 
Radio Frequency Fingerprint Embedding (RFFE) to integrate 
each received signal's bispectrum histogram associated features into the InfoGAN's training process. The work in~\cite{Wang_Comm_2021} adopts a Complex-Valued Neural Network (CVNN) for IoT radio identification using 
complex-valued, baseband signals. It also uses Neural Network (NN) based compression to reduce the dimensionality of the baseband signals for a more computationally efficient CVNN implementation. The work in~\cite{Pan_2019} presents an SEI approach that identifies radios using a 
Deep Residual Network (DRN) and grey-scale images of the received signals' Hilbert spectrum. A received signal's Hilbert spectrum is generated by decomposing each received signal into 
a finite number of intrinsic mode functions via the Hilbert-Huang Transform (HHT). 
The DRN learns discriminatory features from the input grey-scale images to enable identification of one radio over another. The compression system in~\cite{Wang_Comm_2021} and DRN architecture in~\cite{Pan_2019} aim to reduce the complexity of the SEI model, but not the frequency used to sample the baseband signals. These approaches differ from our approach in that none of the reviewed published efforts specifically reduce the SEI hardware complexity by reducing the sampling rate at which an IoT device collects signals for subsequent SEI processing.

\section{Problem Definition
\label{sec:problem_def}}
Traditional IoT networks connect many 
devices and sensors which are assumed to send and receive, and process a few bytes per day at low data rates. This requirement is satisfied by designing the devices to measure and communicate a few values to match the IoT communication technology limits \cite{Krug2018IoTCI}. Satisfying that requirement becomes more challenging for high sampling rate and data intensive applications. The work in~\cite{Wang_Comm_2021,Pan_2019} considers less computationally complex architectures and learning algorithms for a more efficient implementation of SEI in IoT devices.
These efforts did not consider reducing the computational requirement by reducing the sampling rate at which an IoT device collects signals for subsequent SEI.

\section{Objective
\label{sec:obj}}
The goal of this work is to study SEI performance when an IoT device's sampling rate is reduced to decrease computational complexity. This work presents a DL-based approach that leverages a conditional GAN architecture to improve SEI performance at lower sampling rates. Our contributions to the current level of SEI understanding are as follows:
\begin{itemize}
\item{Using adversarial training provided by conditional GAN architecture to upsample IoT waveform while preserving/improving SEI discriminative features.}
\item{Analyzing the effect of low sampling rate on the performance of SEI to discriminate between IoT devices that communicate using IEEE 802.11a.}
\item{Using CGAN-based architecture to improve the traditional low sampling rate SEI performance by an average of 4\% when sampling rate is reduced by a factor of 8.}
\end{itemize}

\section{Background%
\label{sec:background}}
This section provides the necessary background information on the signal of interest, employed deep learning algorithms as well as two interpolation techniques used to facilitate comparative assessment.
\subsection{Signal of Interest%
\label{sec:signal_of_interest}}
The results presented herein are generated using the IEEE 802.11a Wi-Fi preambles from our work in~\cite{Fadul_Access_2021}. The use of IEEE 802.11a Wi-Fi signals is due to (i) its designation as a IoT communications standard \cite{Porkodi_IntelComp_2014}, (ii) its use in other SEI work~\cite{Restuccia_DeepRadioID_2019,Fadul_Access_2021,GlobeCom_Tyler_2021}, and (iii) the fact that current and future communication systems are built upon the Orthogonal Frequency Division Multiplexing (OFDM) scheme~\cite{Lajos_2011}. The data set consists of 8,000 IEEE 802.11a Wi-Fi preambles collected from four Cisco AIR-CB21G-A-K9 radios using an Agilent spectrum analyzer at a sampling rate of 20~{MHz}.
\subsection{Deep Learning Architectures} This section provides a brief description of each DL architecture used in this work.
\subsubsection{Autoencoder (AE)}An AE is a deep generative model that attempts to reconstruct the input data at its output layer \cite{Goodfellow-2016}. An AE consists of an input, hidden, and output layer. Functionally, its comprised of two parts: an encoder and decoder. The encoder generates a representation (a.k.a., a code) of the input data.
The encoder is similar to a Convolutional Neural Network (CNN), but without the fully connected layers. The decoder reconstructs the input using the encoder generated code~\cite{Goodfellow-2016,radar}. An AE does not perfectly copy the input data at its output, but learns how to approximately recover training-like input data under certain restrictions. This allows an AE to learn a good representation and useful properties of the input data~\cite{Goodfellow-2016}. When compared to traditional Multi-Layer Perceptron (MLP) networks, AEs learn a compressed representation from unlabeled, input data using an unsupervised scheme~\cite{DL_Practitioners, Masci_CAE}.

This work uses a Convolutional AE (CAE) for its ability to handle multi-dimensional data such as the complex-valued signals that result from analog-to-digital conversion within the signal collection process. The collected signal's In-phase and Quadrature (IQ) values are restructured to form the first and second rows of a multi-dimensional input, respectively. The CAE's encoder consists of convolutional layers that extract the features and pooling layers to reduce the dimensionality of the resulting feature maps~\cite{radar}. The hidden 
layer is constrained to be smaller than that of the input layer, thus forcing the CAE to learn the most efficient representation of the input data~\cite{Goodfellow-2016}.  Its decoder upsamples and reconstructs the input using de-convolutional and un-pooling layers and the code from the hidden layer.
\subsubsection{Convolutional Neural Networks (CNN)}A CNN is constructed by prepending a feed-forward MLP network with one or more convolutional and pooling layers, which extract and resize multi-dimensional data such as images, or in our case, the IQ values associated with sampled signals~\cite{DL_Practitioners,Riyaz_CNN_2018}. Each convolutional layer generates a feature map from its input, and an activation function performs a linear/nonlinear transformation for each node in the feature map. This work uses the Rectified Linear Unit (ReLU) as the activation function. The activated feature maps are passed to the pooling layer for dimensionality reduction. After one or more convolutional, activation, and pooling layers, dense layers extract higher-level features from the feature maps \cite{Restuccia_DeepRadioID_2019}. Finally, the the output layer assigns the learned features to one of the classes that represent each of the four radios.
\subsubsection{Generative Adversarial Network (GAN)}A GAN uses unsupervised or semi-supervised adversarial training to learn deep representations of the training data ~\cite{Goodfellow_NIPS2014,GAN_overview_2018}. A GAN learns these representations by jointly training two deep networks: the generator, $G$, and discriminator, $D$. The $G$ learns the training data distribution to generate new data samples that minimize the $D$'s probability of correctly determining the origin of a sample. The $G$ can be represented by any deep generative network such as a CAE. The $D$ aims to maximize the probability of correctly determining whether or not an input sample came from the training data or was produced by the $G$~\cite{Goodfellow_NIPS2014}. The $D$ can be implemented using any discriminative network such as a CNN. If MLP-like networks are used for both the $G$ and the $D$, then the entire system can be trained using back-propagation~\cite{Goodfellow_NIPS2014,GAN_overview_2018}.

If the input to the $G$ is $z$ with prior probability $P_{z}(z)$, then the function that maps the input to the output is given by $G(z;~\theta_{g})$, with $\theta_{g}$ being the parameters of $G$. The $D$'s function is given as $D(x;~\theta_{d})$, and its output is a single value representing the probability that the input $x$ is from the training data rather than the $G$~\cite{Goodfellow_NIPS2014}. The GAN's minimax optimization problem is represented by the objective function given by,
\begin{align}
	\underset{G}{\min} \ \underset{D}{\max}~V(D, G) &= E_{x\sim P_{\text{d}}(x)}\{\log[D(x)]\} \nonumber \\
	&+ E_{z\sim P_{z}(z)}\{\log[1 - D(G(z))]\}.
	\label{eqn:GAN_LOSS}
\end{align}
where $E$ is the expected value. During training, the optimum point is reached when the $G$ perfectly recovers the training data distribution (i.e., $P_{g}$$=$$P_{\text{d}}$). In this work, $z$ is IEEE 802.11a Wi-Fi preambles that have been collected at a lower sampling frequency than desired for SEI processing (e.g., 5~MHz versus 20~MHz). The goal is to train the $G$ to recover the higher frequency sampled preamble---from its lower sampled version---while preserving the unique SEI features that permit discrimination of a given radio from the set of known radios.
\subsection{Interpolation Techniques%
\label{sec:interp}}
Interpolation techniques can be used to estimate function values for undefined points within intervals bounded by points with defined values (a.k.a., measurements). Interpolation can be employed to upsample the collected IoT signals since they are composed of a sequence of points in time paired with sample values (a.k.a., measurements). In this work, the sampling frequency--of the IoT devices' signals--is performed using piece-wise Linear Approximation Interpolation (LAI) and Cubic-Spline Interpolation (CSI). The higher sampling frequency is expressed as,
\begin{equation}
    F_{H} = V \times F_{L},
\end{equation}
where $V$ is an integer associated with the increase applied to the lower sampling frequency $F_{L}$.
%

In piece-wise linear approximation, if the signal to be interpolated $z$ is defined at some points $a = \tau_{1} < \tau_{2} < \cdots < \tau_{n} = b$, then the interpolating function $f$ can be represented by pieces of linear functions bounded by the intervals $[\tau_{k} , \tau_{k+1}]$~\cite{deboor}. On each interval $[\tau_{k}, \tau_{k+1}]$, the function $f$ and signal $z$ are exactly equal at the interval points, while $f$ approximates $z$ with a straight line connecting $\tau_{k}$ and $\tau_{k+1}$ within the interval~\cite{deboor}.     

In CSI, if the function values $z(\tau_{1})$, $z(\tau_{2})$, $\cdots$, $z(\tau_{n})$ are defined for the interval points $\tau_{1} < \tau_{2} < \cdots < \tau_{n}$, the points between interval points can be estimated using a piece-wise cubic interpolating function $f$, where each piece $P_{i}$ is at most a third-degree polynomial \cite{deboor}. The pieces $P_{i}$ of the cubic-spline interpolating function $f$ satisfies the following conditions:
\begin{align}
P_{i}(\tau_{i}) &= z(\tau_{i}), \\ P_{i}(\tau_{i+1}) &= z(\tau_{i+1}), \\
P_{i}^{\prime}(\tau_{i}) &= s_{i}, \\ P_{i}^{\prime}(\tau_{i+1}) &= s_{i+1},
\label{eqn:spline}
\end{align}
where $i=1, \cdots, n-1$, and $s_{1}, \cdots, s_{n}$ are free parameters \cite{deboor}. The conditions in equation~\eqref{eqn:spline} state that the cubic-spline interpolating function $f$ must be continuous and have a continuous first derivative over the entire interval $[a, b]$.
\begin{table}[!t]
\processtable{Generator, $G$, NN configurations for signal collection sampling frequencies of $F_{L}$$=$$[2.5, 5, 10]$~MHz.}
{\includegraphics[width=\columnwidth]{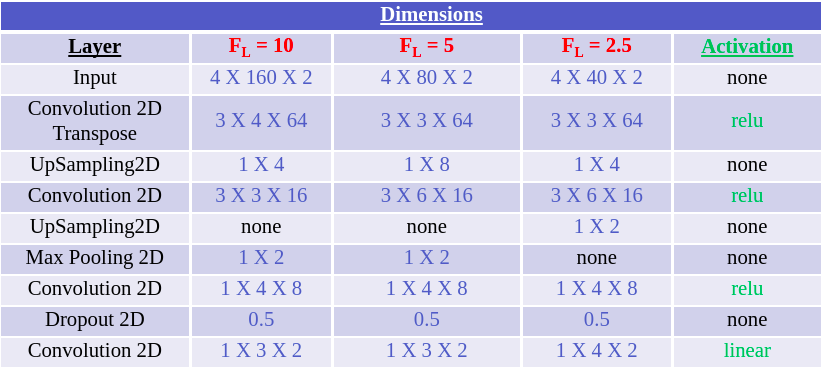}}{\label{tab:G_arch}}
\vspace{-5mm}
\end{table}
\section{Methodology%
\label{sec:methodology}}
\subsection{Data Preprocessing%
\label{sec:data_proc}}
\begin{figure}[!b]
	\centering
	\vspace{-5mm}
	\includegraphics[width=\columnwidth]{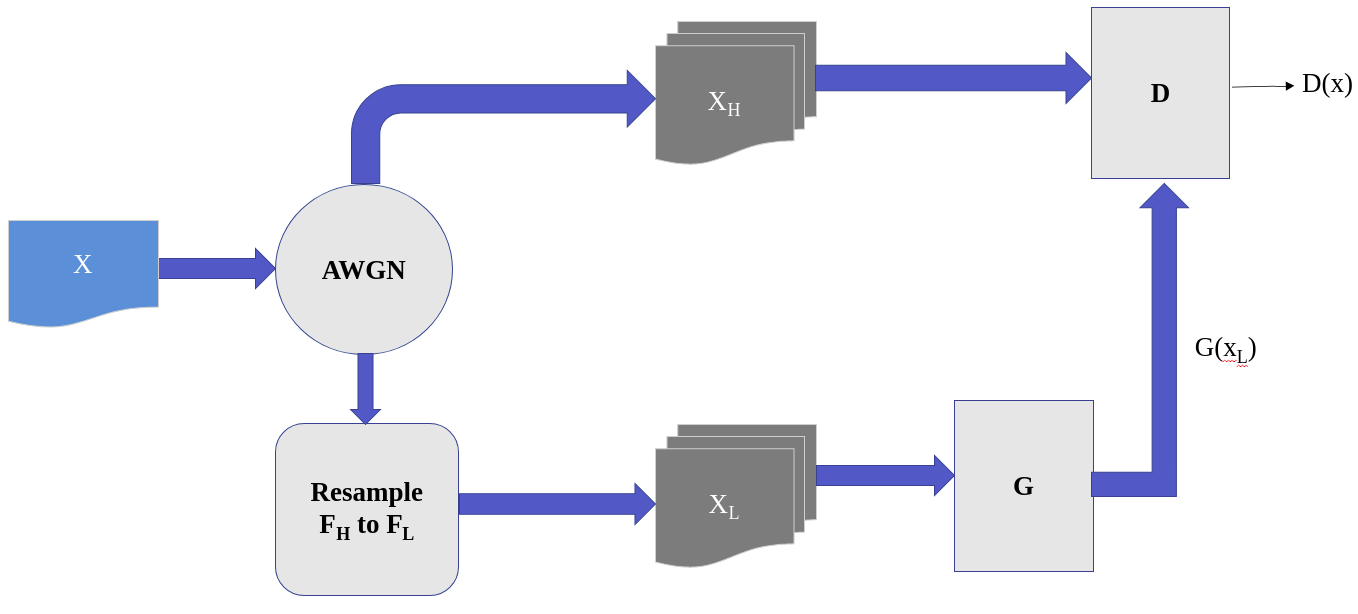}
	\caption{Flowchart illustrating the process used to preprocess the collected signal data and train the cGAN to facilitate upsampling while preserving radio-specific SEI features.}
	\label{fig:GAN_training}
\end{figure}

Fig.~\ref{fig:GAN_training} shows the overall process developed for data preprocessing and cGAN training. The cGAN is formed using a CAE and CNN for the $G$ and $D$ networks, respectively. A total of 2,000 preambles, $X$, are collected from each of the four Wi-Fi radios. Due to the $F_{H}$$=$$20$~{MHz} sampling rate, each preamble consists of 320 IQ samples and has like-filtered Additive White Gaussian Noise (AWGN) added to it to achieve a Signal-to-Noise Ratio (SNR) of 9~{dB} to 30~{dB} in increments of 3~{dB} between consecutive values. This process is repeated ten times per preamble to augment the training set and facilitate Monte Carlo analysis. For a given SNR and each radio, 1,600 preambles from each noise realization are randomly selected to form the training data set, $X_{H}$. The training data is copied and downsampled to one of three lower sampling rates of $F_{L}$$=$$[2.5, 5, 10]$~{MHz} to form the $G$ inputs, $X_{L}$. The preambles not chosen for the training data set are used to generate the results. All of the preambles are formatted in accordance with the procedure described in~\cite{GlobeCom_Tyler_2021}, which forms a two-dimensional (2D) tensor constructed from a preamble's IQ samples and their natural logarithm. Thus, the cGAN is learning the deep representation from the preambles' raw IQ samples, magnitude, and phase representations. Each 2D tensor is scaled column-wise using min-max normalization.
\subsection{Deep Networks Construction, Training, and Usage}
In cGAN, the mapping functions $G(z)$ and $D(x)$ are conditioned by the class label $y$; thus, the minimax optimization equation \eqref{eqn:GAN_LOSS} is rewritten as,
\begin{align}
	\underset{G}{\min} \ \underset{D}{\max}~V(D, G) &= E_{x\sim P_{\text{d}}(x)}\{\log[D(x|y)]\} \nonumber \\
	&+ E_{z\sim P_{z}(z)}\{\log[1 - D(G(z|y))]\},
	\label{eqn:cGAN_LOSS}
\end{align}
where $P_{\text{d}}(x)$ is the training data distribution learned by the $G$. The class label $y$ is added to the $G$ and $D$ inputs using a hidden representation. This enables the cGAN to estimate a one-to-many generative function instead of the traditional one-to-one mapping. The hidden representation is generated using an 
embedding layer to generate a length 50 label vector for each of the four Wi-Fi radios. The resulting vectors are expanded using a Fully Connected (FC) layer with a linear activation function, so that they can be reshaped and added---as an extra channel---to each preamble's 2D representation. 
For $X_{H}$, the output of the FC layer is of size 1,280, reshaped into a 4$\times$320 matrix, and appended to each preamble's 2D tensor along the third dimension to form a 4$\times$320$\times$2 representation. The only difference between $X_{L}$ and $X_{H}$ is the number of columns comprising the final three-dimensional tensor. The number of columns depends on the lower sampling rate being investigated. For lower sampling rates of [2.5, 5, and 10]~{MHz} the number of columns is 40, 80, and 160, respectively. 
The NN architectures associated with the $G$ and $D$ are provided in Table~\ref{tab:G_arch} and Table~\ref{tab:D_arch}, respectively. 
The $F_{L}$$=$$2.5$~{MHz} case's $G$ is constructed using an additional upsampling layer in lieu of the max pooling layer---used in the other cases---to ensure its output dimension is consistent with the $F_{H}$$=$$20$~MHz case (i.e., $4$$\times$$320$$\times$$2$).
\begin{table}[!t]
\processtable{Configurations used to construct the discriminator, $D$, NN and CNN used for SEI.}
{\includegraphics[width=\columnwidth]{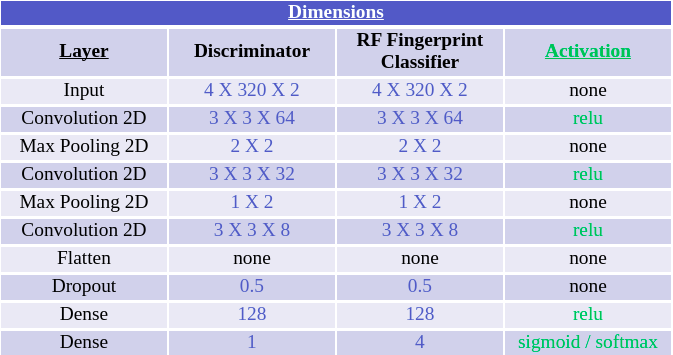}}{\label{tab:D_arch}}
\vspace{-5mm}
\end{table}
The cGAN is trained using backpropagation with a 256 tensor mini-batch, 
1,000 epochs, and an alternating scheme in which the $D$ is trained for a given $G$ in $k$ steps. The latter results in the best $D$ for that particular $G$. The work in~\cite{Goodfellow_NIPS2014} treats $k$ as a hyperparameter with a choice of one being the least computationally complex; thus, we set the value of $k$ to one. The $D$'s training utilizes forward- and backpropagation with the goal of maximizing $V(D, G)$ to achieve the highest probability for making a correct decision. After $k$ steps, the $G$ is trained using stochastic gradient descend to minimize the function $V(D, G)$, with the goal of working against the ability of the $D$ to make a correct decision. This training process continues until $D(x)$$=$$0.5$ (i.e., the discriminator is guessing as to the origin of the input) everywhere or the total number of training iterations equals the empirically chosen value of 1,000.

\begin{figure}[!b]
	\centering
	\vspace{-5mm}
	\includegraphics[width=\columnwidth]{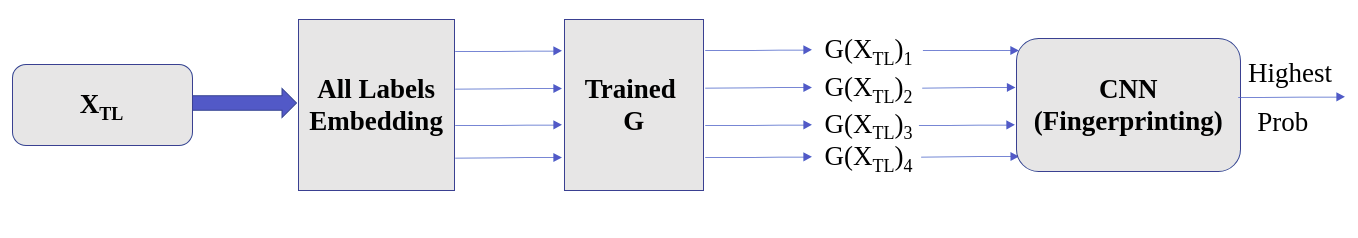}
	\caption{Flowchart illustrating the process used to preprocess the collected signal data and train the cGAN to facilitate upsampling while preserving radio-specific SEI features.}
	\label{fig:GAN_fingerprinting}
\end{figure}
\indent Once cGAN training concludes for the selected sampling rate $F_{L}$, the resulting $G$---that provides the mapping function $G(z, \theta_{g})$---is disconnected from the $D$ and used to upsample preambles collected at rate $F_{L}$ to the desired sampling rate of $F_{H}$. The second channel---introduced for cGAN training---is removed from the resulting upsampled preambles prior to SEI using a trained CNN. This process is illustrated in Fig.~\ref{fig:GAN_fingerprinting}. The CNN is trained using (i) the 2D tensors that form the training set $X_{H}$, (ii) backpropagation, (iii) stochastic gradient descent optimization to minimize the categorical cross-entropy loss function, (iv) $l_{2}$ regularization to reduce overfitting, and (v) adjustment of the network's weights via Adam optimization~\cite{adam_optimizer}. The Adam optimizer's initial learn rate is set to $1$$\times$$10^{-3}$. The final layer of the CNN is a softmax activation that assigns the $i^{\text{th}}$ preamble a label of $Q_{i}$ according to
\begin{equation}
	Q_{i} = \max_{j}(y_{ij}),
	\label{eqn:softmax}
\end{equation}
where $j$$=$$[1,2, 3, 4]$ and $i$=1, 2, $\dots$, 1,600 for a given radio, SNR, and noise realization. 

{Best possible SEI performance is achieved by training the CNN using preambles that are at an SNR equal to or lower than those comprising the test set. A grid search determined that training at SNR values of [9, 9, 9, 12, 15, 15, 15, 18]~{dB} resulted in superior SEI performance when classifying test preambles at SNR values of [9, 12, 15, 18, 21, 24, 27, 30]~{dB}, respectively.}
\section{Results%
\label{sec:results}}
\begin{figure}[!t]
	\centering
	\includegraphics[width=\columnwidth]{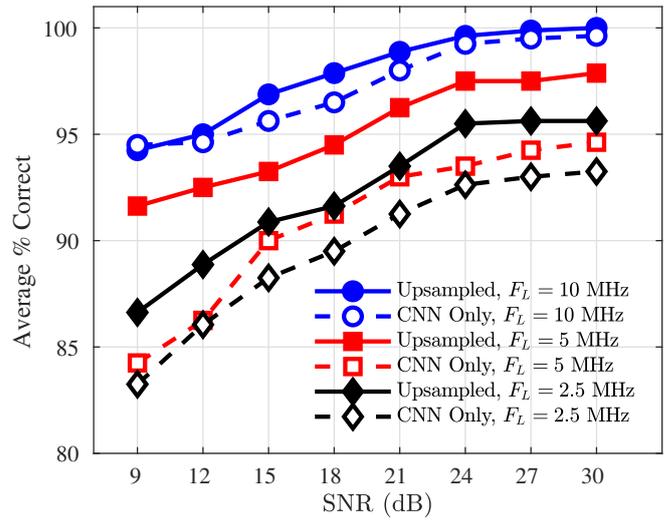} 
	\caption{Average percent correct classification performance results generated from IEEE 802.lla Wi-Fi preambles that are transmitted by four radios, collected at sampling rates of $F_{L}$ equal to 2.5~{MHz}~$(\Diamond)$, 5~{MHz}~$(\square)$, or 10~{MHz}~$(\bigcirc)$, and upsampled to a frequency of 20~{MHz} using a cGAN prior to CNN classification (designated using solid lines and filled markers). The dashed lines and unfilled markers designate average percent correct classification results generated using the ``CNN only'' case in which training and testing is conducted using preambles collected at the sample frequencies $F_{L}$ (i.e., cGAN-based upsampling is not conducted).}
	\label{fig:Upsampling_Res}
	\vspace{-5mm}
\end{figure}

For lower sampling frequencies $F_{L}$$=$$[2.5, 5, 10]$~{MHz}, Fig.~\ref{fig:Upsampling_Res} shows the average percent correct classification performance for the upsampled IEEE 802.11a Wi-Fi preamble tensors generated using the corresponding trained $G$ and CNN at SNR values of 9~{dB} to 30~{dB} in 3~{dB} steps. These results are presented using solid lines and closed markers. In addition to these results, average percent correct classification performance is shown for the case when a CNN is trained and tested using tensors constructed directly from preambles sampled at the lower frequency $F_{L}$. These results are designated as the ``CNN only'' cases, included for comparative assessment, and presented using dashed lines and open markers. All of the presented results are generated using Tensorflow 2.0 running on NVIDIA Tesla K40m Graphics Processing Units (GPUs).

\begin{figure}[!b]
	\centering
	\vspace{-5mm}
	\includegraphics[width=\columnwidth]{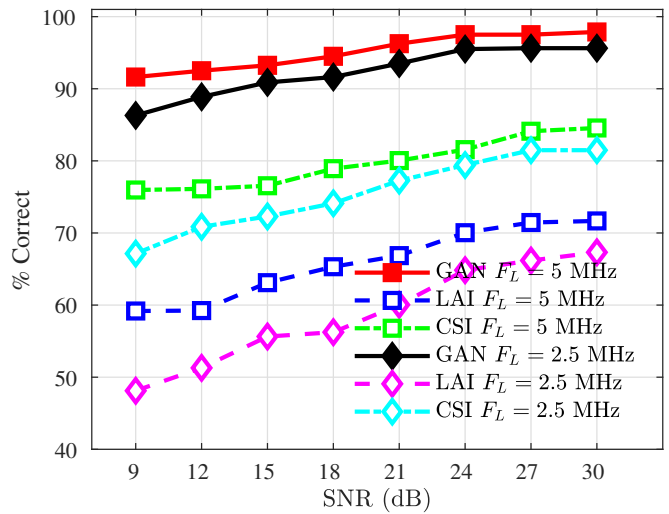} 
	\caption{Average percent correct classification performance results generated from IEEE 802.lla Wi-Fi preambles that are transmitted by four radios, collected at sampling rates of $F_{L}$ equal to 2.5~{MHz}~$(\Diamond)$ and 5~{MHz}~$(\square)$ and upsampled to a frequency of 20~{MHz} using a cGAN (solid lines and closed markers). Comparative assessment is facilitated by including the CNN-based SEI results corresponding with the LAI (dashed lines and unfilled markers) and CSI (dashed-dot lines and unfilled markers) upsampling approaches.}
	\label{fig:int}
\end{figure}
The presented cGAN-driven upsampling approach results in superior SEI performance over the ``CNN only'' approach for SNR values equal to or greater than 9~{dB} when $F_{L}$ is equal to 2.5~{MHz} or 5~{MHz} and 12~{dB} for $F_{L}$ equal to 10~{MHz}. When considering an average percent correct performance threshold of 90\%, the ``CNN only'' case achieves or exceeds this threshold for all SNR values when using preambles sampled at a frequency $F_{L}$$=$$10$~{MHz}. However when using preambles sampled at rates of $F_{L}$$=$$5$~{MHz} or $F_{L}$$=$$2.5$~{MHz} the ``CNN only'' case meets or exceeds the threshold at SNR values greater than or equal to 15~{dB} and 21~{dB}, respectively. In contrast, when using preambles that are upsampled---using the corresponding trained $G$ network---SEI performance meets or exceeds the 90\% threshold for the lower sampling frequencies, $F_{L}$, of 5~{MHz} and 10~{MHz} at SNR values of 9~{dB} and higher. When the trained $G$ is used to upsample preambles collected using a sampling frequency of $F_{L}$$=$$2.5$~{MHz}, then the 90\% threshold is met or exceeded at SNR values of 15~{dB} and higher. Interestingly, the greatest SEI performance improvement occurs when the radios' preambles are collected uing a sampling frequency of $F_{L}$$=$$5$~{MHz} and upsampled to $F_{H}$$=$$20$~{MHz} using the $G$ from the corresponding trained cGAN. For this case, the largest improvement in SEI performance is roughly 7\% (i.e., 84\% to 91\% when comparing ``CNN only'' to our approach) at an SNR of 9~{dB}. While the smallest improvement is 3\% at an SNR value equal to 27~{dB}. This is surprising as it seems reasonable to expect that the largest percentage improvement would occur when the $F_{L}$$=$$10$~{MHz} sampled preambles are upsampled to 20~{MHz} prior to CNN classification. This expectation is based upon the fact that higher sampling frequencies capture more of the radio-specific features exploited by the SEI process, so there are fewer of these features that need to be learned and ``filled in'' by the $G$ for the 10~MHz upsampling case versus the other two. The SEI performance improvement observed for preambles that are upsampled to $F_{H}$$=$$20$~{MHz} from $F_{L}$$=$$5$~{MHz} is \textit{not} achieved when using signals sampled at a frequency of $F_{L}$$=$$2.5$~{MHz} during collection. This indicates that there is a lower limit to the presented approach.

\begin{figure*}[!t]
\begin{minipage}[t]{0.5\textwidth}
	\centering
	\subfigure[Collected sampling frequency of $F_{L}$$=$$5$~{MHz}.]{\label{fig:Individual_5}\vspace{2in}
	\includegraphics[width=0.9\columnwidth]{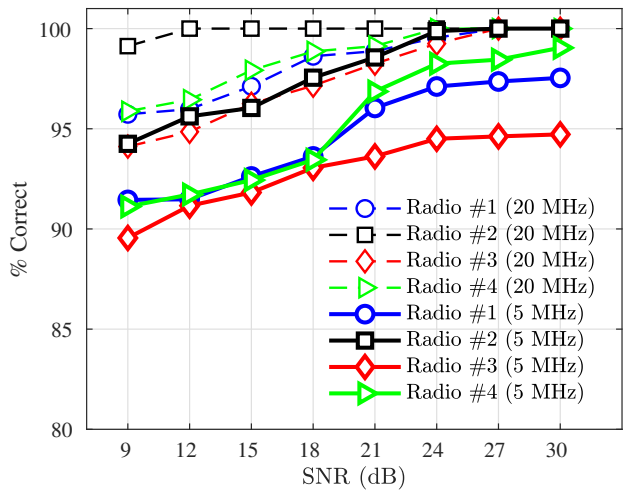}}
\end{minipage}
\begin{minipage}[b]{0.5\textwidth}
    \centering
	\subfigure[Collected sampling frequency of $F_{L}$$=$$2.5$~{MHz}.] {\label{fig:Individual_2}
	\includegraphics[width=0.9\columnwidth]{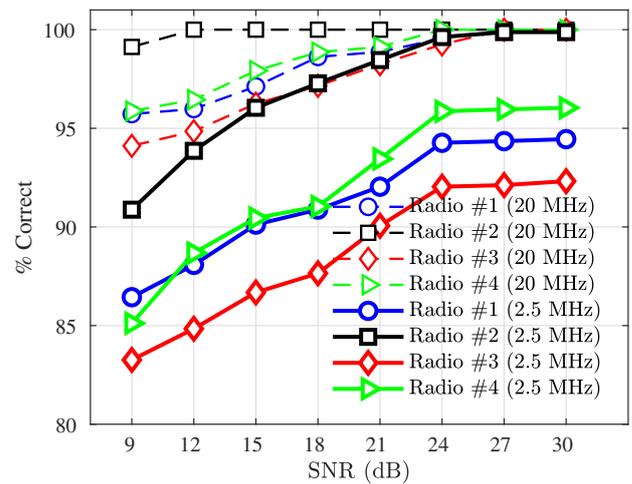}}
\end{minipage}
\caption[]{Percent correct classification performance results (solid lines) generated from IEEE 802.lla Wi-Fi preambles that are transmitted by four radios, collected at sampling frequencies of $F_{L}$ equal to 2.5~{MHz} or 5~{MHz}, and upsampled to a frequency of $F_{H}$$=$$20$~{MHz} using a cGAN prior to CNN classification. The percent correct classification results associated with the CNN classification of these same preambles--but collected using a sampling frequency of 20~{MHz}--are included for completeness and illustrated using dashed lines.}
	\vspace{-5mm}
	\label{fig:Individual}	
\end{figure*}

\begin{figure}[!t]
\begin{minipage}[t]{\columnwidth}
	\centering
	\subfigure[Collected sampling frequency of $F_{L}$$=$$10$~{MHz}.]{\label{fig:Upsampling_Soa_10}\vspace{2in}
	\includegraphics[width=0.95\columnwidth]{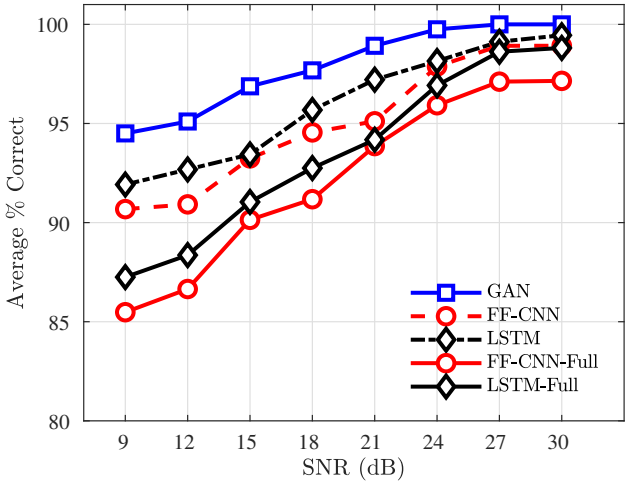}}
\end{minipage}
\begin{minipage}[t]{\columnwidth}
	\centering
	\subfigure[Collected sampling frequency of $F_{L}$$=$$5$~{MHz}.]{\label{fig:Upsampling_Soa_5}\vspace{2in}
	\includegraphics[width=0.95\columnwidth]{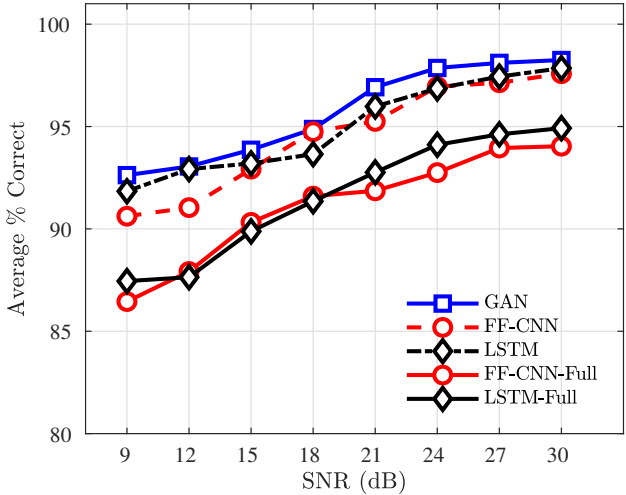}}
\end{minipage}
\begin{minipage}[b]{\columnwidth}
    \centering
	\subfigure[Collected sampling frequency of $F_{L}$$=$$2.5$~{MHz}.] {\label{fig:Upsampling_Soa_2}
	\includegraphics[width=0.95\columnwidth]{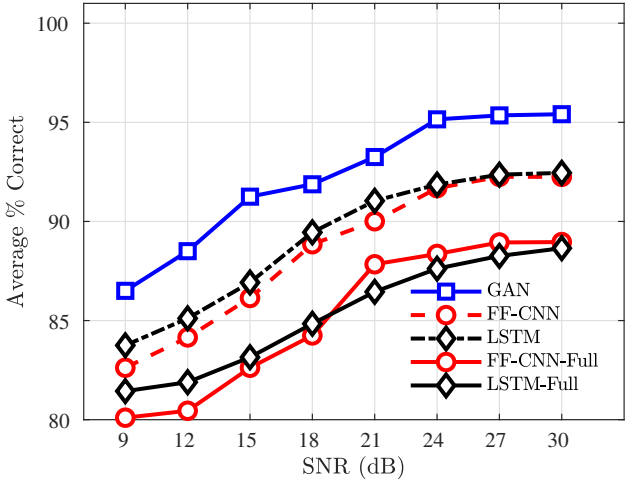}}
\end{minipage}
\caption[]{Average percent correct classification performance results generated from IEEE 802.lla Wi-Fi preambles that are transmitted by four radios, collected at sampling rates of $F_{L}$ equal to 2.5~{MHz}, 5{MHz}, or 10~{MHz}, and (i) upsampled to a frequency of 20~{MHz} using a cGAN prior to CNN classification ($\square$), (ii) Classified using variable-length Flatten-Free CNN (FF-CNN) ($\bigcirc$) and LSTM ($\Diamond$) in~\cite{Shen_Versatile_2022}. The Average percent correct classification performance results designated with "Full" are the ones that include $F = 20$~{MHz} preambles in the training set.}
	\label{fig:Upsampling_Soa}	
\end{figure}

The presented cGAN-driven upsampling approach (solid lines and filled markers) is compared with the LAI (dashed lines and unfilled markers) and CSI (dash-dot lines and unfilled markers) upsampling approaches--as described in Sect.~\ref{sec:interp}--using average percent correct classification performance for lower sampling frequencies of 2.5~{MHz} ($\Diamond$) and 5~{MHz} ($\square$) for SNR values of 9~{dB} to 30~{dB} in steps of 3~{dB} between consecutive values, Fig.~\ref{fig:int}. Results associated with the $F_{L}$$=$$10$~{MHz} lower sampling frequency case are omitted for the following reasons: (i) only a marginal improvement (i.e., maximum of 3\%) is observed between the cGAN upsampled and ``CNN only'' results shown in Fig.~\ref{fig:Upsampling_Res}, and (ii) their omission improves visual clarity of the 2.5~{MHz} and 5~{MHz} results without diminishing the value of the comparative assessment. For the average percent correct classification results presented in Fig.~\ref{fig:int}, the cGAN-driven upsampling approach results in superior performance over the results corresponding to the LAI and CSI approaches for all investigated SNR values. As a matter of fact the results associated with the LAI and CSI approaches never achieve the 90\% average percent correct performance threshold for any of the investigated SNR values. These results show that the cGAN's ability to learn the radios' SEI distributions is required to upsample signals prior to CNN-based discrimination. The CSI approach provides superior performance over LAI. When using $F_{L}$$=$$2.5$~{MHz} sampled preambles, the CSI approach's average percent correct classification performance outperforms the results associated with the LAI approach using preambles sampled using a frequency of $F_{L}$$=$$5$~{MHz} for all SNR values. The largest margin--between the CSI's $F_{L}$$=$$2.5$~{MHz} results and the LAI's $F_{L}$$=$$5$~{MHz} results--is {11\% (i.e., 70\% versus 59\%) at an SNR value of 12~{dB} and the smallest is 8\% at an SNR of 9~{dB}.} The CSI performance improvement--over the LAI approach--is attributed to its use of a \textit{cubic} interpolating function instead of the linear function used by LAI. This allows the CSI approach to better model the behaviors present in signals collected at a sampling frequency of $F_{H}$$=$$20$~{MHz}. 
%
%
Fig.~\ref{fig:Individual} presents percent correct classification performance results for each radio whose collected IEEE 802.11a Wi-Fi preambles are upsampled to a sampling frequency of $F_{H}$$=$$20$~{Mhz} using a cGAN. For completeness, Fig.~\ref{fig:Individual} also includes the individual radio percent correct classification performance for the case when CNN-based SEI is performed using preambles collected at a sampling frequency of 20~{MHz} (i.e., no upsampling) and are indicated using thinner lines than those associated with the cGAN upsampled results. The results shown in Fig.~\ref{fig:Individual_5} and Fig.~\ref{fig:Individual_2} correspond to IEEE 802.11a Wi-Fi preambles that are collected at sampling frequencies of $F_{L}$$=$$5$~{MHz} and $F_{L}$$=$$2.5$~{MHz}, respectively. The $F_{L}$$=$$5$~{MHz} SEI results, Fig.~\ref{fig:Individual_5}, outperform that of the $F_{L}$$=$$2.5$~{MHz} case, Fig.~\ref{fig:Individual_2}, and experience less spread across the four radios. These results are expected as the cGAN is filling in 87.5\% of the IQ samples--along with their associated SEI features--for the $F_{L}$$=$$2.5$~{MHz} case versus 75\% when the original sampling frequency is $F_{L}$$=$$5$~{MHz}. When considering this information, the presented results are quite impressive considering that individual radio discrimination performance for the $F_{L}$$=$$2.5$~{MHz} and $F_{L}$$=$$5$~{MHz} cases remain above 83\% and 89\% for all radios and SNR values. The relative performance of the four radios remains consistent with Radio \#2 being the most easily distinguished and Radio \#3 being the hardest to distinguish. For both upsampling cases--going from 5~{MHZ} to 20~{MHz} and 2.5~{MHZ} to 20~{MHz}--the percent correct classification performance is poorer than the case of SEI performed using preambles collected at a sampling frequency of 20~{MHz}. This suggests that the cGAN is not completely capturing the distributions of the SEI features present in the four radios' waveforms. Some possible solutions could be to use more waveforms in the training of the cGAN, use of an alternate waveform representation (e.g., Fourier or Gabor transform), or a combination of the two.  
\subsection{Comparison with State-of-Art SEI}
The performance of the SEI approach presented in this paper is compared against the one in~\cite{Shen_Versatile_2022}. The work in~\cite{Shen_Versatile_2022} presents a DL-based SEI protocol to identify wireless devices with the ability to process variable-length NN inputs and use online augmentation to improve the SEI performance at low SNRs. The proposed SEI protocol is tested using LoRa/LoRaWAN technology devices which is a popular low power wide area network (LPWAN) protocol. LoRaWAN adopts an Adaptive Data Rate (ADR) mechanism that enables wireless devices to adjust transmission configuration in real time~\cite{Shen_Versatile_2022}. The ADR capability enables the LoRa devices to adjust the length of the preamble and payload to tackle continuously changing wireless channel conditions. \\
LoRa modulates data at the physical layer using Chirp Spread Spectrum (CSS). Three pieces of information are needed to generate a LoRa preamble using CSS modulation: (i) signal amplitude, (ii) bandwidth, and (iii) Spreading Factor (SF). The SF parameter controls the length of the signal and results in increasing bandwidths from 125~{kHz} to 500~{kHz} for SF values of 7 to 12 as detailed in~\cite{Shen_Versatile_2022}. 
Our selection of the approach in~\cite{Shen_Versatile_2022} 
is due to the fact that both our approach and the one in~\cite{Shen_Versatile_2022} are designed to process variable length inputs (a.k.a., preambles). In~\cite{Shen_Versatile_2022}, the length of the preamble is controlled by the CSS modulation's SF parameter, while in our approach the preamble length is controlled by the sampling frequency. 
\\
The work in~\cite{Shen_Versatile_2022} represents LoRa preambles using a channel independent spectrogram to mitigate the wireless channel effects. Each preamble's spectrogram $S^{i}$ 
is 
a matrix of $N \times (M-1)$ elements with each element designated by $S^{i}_{k,m}$ in which $k=1,2,\dots,N$ and $m=1, \dots, M - 1$. $N$ represents the length of the spectrogram window, which is set to sixty-four 
in~\cite{Shen_Versatile_2022}. $(M-1)$ is the width of $S^{i}$ where $M$ is related to $N$ by,
\begin{equation}
    M = \dfrac{8 \times \left(\dfrac{2^{SF}}{B}\right) \times F_{L} - N }{R} + 1,
\end{equation}
where $F_{L}$ is the sampling frequency (a.k.a., the reciprocal of the sampling interval), and $R$ is the hop size which is always set to thirty-two by the authors of~\cite{Shen_Versatile_2022}. In \cite{Shen_Versatile_2022}, the spectrogram $S^{i}$ has a fixed length of $N=64$, $F_{s} = 250$~{KHz}, and a variable width of $M-1$ that depends on the selected spreading factor $SF$. 
In order to compare our approach with that in~\cite{Shen_Versatile_2022}, we generate spectrograms in which the width, $M-1$, is determined by the sampling frequency, $F_{L}$, instead of the spreading factor $SF$. So, in our case the spreading factor is fixed and the sampling frequency changes, which is the opposite of~\cite{Shen_Versatile_2022}. 
For the results presented in this section, spectrograms $S^{i}$ are generated for all IEEE 802.11a preambles--described in Sect.~\ref{sec:signal_of_interest}--for $F_{s} = [2.5, 5, 10]$~{MHz} and $SF=7$ for all sampling frequencies.

The authors of~\cite{Shen_Versatile_2022} introduce online augmentation to increase the SEI system's robustness to noise. In online augmentation, AWGN is added to each minibatch--selected from the training set--directly before being passed to the NN. 
In online augmentation, the NN model is trained using (steps $\times$ minibatch size $\times$ noise realizations) noisy signals. The results in this section are generated using online augmentation with 13 steps, a minibatch size of 128, and ten noise realizations. The authors of~\cite{Shen_Versatile_2022} use four NN architectures to process the variable-width, channel independent spectrograms generated from the variable length, LoRa preambles. Based on the results presented in~\cite{Shen_Versatile_2022}, all NN architectures exhibit similar performance. Therefore, only two NNs are selected for the comparative assessment, which are a: 
(i) Flatten-Free CNN (FF-CNN) where the flatten layer of a ResNet network is replaced with a global average pooling 2D layer that always provides a fixed-length vector to the dense layers regardless of the spectrogram's width at the input of the FF-CNN, and (ii) LSTM, where a global average pooling 1D layer is used in a similar way as in the FF-CNN to produce a fixed-length vector for the following dense layers. Detailed descriptions of the FF-CNN and LSTM architectures are presented in~\cite{Shen_Versatile_2022}. 

Comparative assessment is enabled by regenerating the results shown in Fig.~\ref{fig:Upsampling_Res}, but with online augmentation integrated into the GAN training process. Average percent correct classification performance is shown for the: regenerated GAN-based upsampling as well as the FF-CNN and LSTM results using IEEE 802.11a Wi-Fi spectrograms for all three sampling frequencies when the classification model is trained with and without the original 20~{MHz} sampled preambles included. 
It can be seen that GAN-based SEI average percent correct classification performance exceeds that of the FF-CNN and LSTM spectrogram-based approaches for all SNR values for sampling frequencies of 10~{MHz}, Fig.~\ref{fig:Upsampling_Soa_10}, and 2.5~{MHz}, Fig.~\ref{fig:Upsampling_Soa_2}. For $F_{L} = 5$~{MHz}, Fig.~\ref{fig:Upsampling_Soa_5}, all three SEI approaches exhibit the similar performance for SNR values of 15~{dB} and above; however, GAN- and LSTM-based SEI average percent correct classification performance surpasses that of the FF-CNN for SNR values below 15~{dB}. 

The SEI approach in~\cite{Shen_Versatile_2022} allows identification of emitters using signals of variable-length by learning length independent features. Our observation is based on the fact that all of the signals--regardless of length--are mapped to the same length dense layer for both NNs (a.k.a., FF-CNN and LSTM); thus, the sampling frequency of the signal to be classified does not change. However, this is problematic when length-variability is due to changing sample frequencies, because the signal points measured during the continuous to discrete conversion change. Hence, the SEI features present in the discrete signals change as well, which leads to feature inconsistencies across different length signals that subsequently inhibits learning of emitter discriminating features. Our cGAN approach does not suffer from these limitations, because it is trained using the 20~{MHz} sampled signals such that its $G$ uses the learned, SEI feature distributions to map the lower sampling rate signals' SEI features into the 20~{MHz} sampled signals' space. In other words, the $G$ is constructing 20~{MHz} sampled versions of the $F_{L}$ sampled signals (a.k.a., upsampling). Thus, in our approach the trained CNN performs SEI using only 20~{MHz} sampled signals and not lower sampling frequency signals that have been ``flattened'' to meet the same, fixed length. cGAN approach performance does degrade as $F_{L}$ decreases. However, this is attributed to the limited size of the 20~{MHz} sampled data set (i.e., only 2,000 preambles per emitter), because the limited data set size impedes learning of emitter accurate SEI feature distributions.
%
%
%
\section{Conclusion%
\label{sec:conclusio}}
This work presents a novel DL-based approach that uses a trained cGAN's generative network to upsample the collected preambles of four IEEE 802.11a Wi-Fi radios from sampling frequencies of [2.5, 5, and 10]~{MHz} to 20~{MHz} in an effort to improve SEI performance while reducing IoT device hardware requirements that collect the signals of the radio(s) being identified. The cGAN-based approach is compared with two traditional interpolation approaches as well as an SEI approach in which the collected signals are \textit{not} upsampled prior to CNN-based classification. The results show that the cGAN-based approach is superior for all three lower sampling frequencies. The largest improvement in average percent correct classification performance occurs when 5~{MHz} sampled signals are upsampled to 20~{MHz}. Our investigation shows that a trained DL architecture can be used to upsample signals while learning a given radio's SEI features so that the SEI process' discrimination performance increases. Future research will focus on improving the SEI performance of the upsampled signals as well as increasing the number of radios used.

\balance

\typeout{}
\bibliographystyle{IEEEtran}
\bibliography{iet_jrnl_of_eng_bib__v01.bib}

\end{document}